\documentclass[12pt]{iopart}

\usepackage{iopams}
\usepackage{graphicx}
\usepackage[sort&compress,numbers,square,comma]{natbib}

\usepackage{color}
\definecolor{orange}{rgb}{1.0,0.3,0.0}

\begin{document}

\paper[Electron emission in collisions between He and Li$^{\;q+}$ ions]%
{Experimental and theoretical results on electron emission in collisions between He targets and dressed Li$^{q+}$ ($q=1,2$) projectiles}
\author{D.~Fregenal$^\ddag$,
 J.~M.~Monti$^\dag$,
 J.~Fiol$^\ddag$,
 P.~D.~Fainstein$^\ddag$,
 R.~D.~Rivarola$^\dag$,
 G.~Bernardi$^\ddag$,
 S.~Su\'{a}rez$^\ddag$
 }
\address{$^{\dag}$\ Laboratorio de Colisiones At\'{o}micas.  Instituto de F\'{\i}sica Rosario (CONICET-UNR) and Facultad de Ciencias Exactas, Ingenier\'{\i}a y Agrimensura, Universidad Nacional de Rosario, Avenida Pellegrini 250, 2000 Rosario, Argentina}
\address{$^{\ddag}$\ Centro At\'{o}mico Bariloche, Comisi\'{o}n Nacional de Energ\'{\i}a At\'{o}mica, 8400 San Carlos de Bariloche (R\'{\i}o Negro) and Consejo Nacional de Investigaciones Cient\'{\i}ficas y T\'{e}cnicas (CONICET), Argentina.}
\eads{\mailto{fregenal@cab.cnea.gov.ar}}

\begin{abstract}
  We investigate experimentally and theoretically the electron emission in collisions between He atoms and $\mathrm{Li}^{q+}$ ($q=1,2$) projectiles at intermediate-high incident energies.  We report on measured absolute values of double differential cross-sections, as a function of the emitted electron energy and angle, at a collision energy of 440~keV/u.  The different contributions from target-ionisation, projectile-ionisation, and simultaneous target-projectile ionisation are calculated with the quantum-mechanical Continuum Distorted Wave and Continuum Distorted Wave -- Eikonal Initial State models, and with Classical Trajectory Monte Carlo simulations.  There is an overall good agreement of the calculations with the experimental data for electron emission cross-sections.
\end{abstract}

\submitto{\jpb} 

\maketitle

\section{Introduction}

The study of electronic reactions in collisions between partially-dressed projectiles and atomic and molecular targets has received an increasing interest in the last two decades.  These systems provide a suitable frame to investigate the relative importance of the electron-electron versus the electron-nucleus interactions.  Particularly, electron emission has received great part of the attention given that it is the main mechanism leading to energy loss of swift ions in matter.  Therefore, detailed knowledge of the mechanisms of electron emission in collisions between fast partially-dressed ions with atomic and molecular targets is relevant in many scientific areas, such as astrophysics, plasma physics, and it also
plays a predominant role in applied areas like the design of fusion reactors, radiation damage and hadron therapy.
A large amount of experimental data for doubly-differential cross-sections (DDCS) are available (see for example the appendix~C of~\cite{Stolter1997_EEI}) involving bare and partially-dressed ions impinging on atomic or molecular targets at intermediate and high collision energies.
However, to the best of our knowledge, besides the results presented by Monti {\it et al.}~\cite{Monti2012JPBp5202}, where the ionisation of He atoms by Li$^{3+}$ impact was investigated, there are no reports of DDCS for electron emission considering Li ions as projectiles nor as targets.

In this work we compare the experimental data with well-known theories, that have been tested for many different systems: a four-body Classical Trajectory Monte Carlo (CTMC) model, and two quantum-mechanical modified distorted-wave models, reported previously by Monti \etal~\cite{Monti2008JPBp201001,Monti2011JPBp195206}.

CTMC has been shown to be a very versatile technique to study collisions involving many bodies.%
The main difficulties with this approach arise from the well-known unstabilities of many-electron atoms, that lead to spontaneous break-up. %
During the last decades, several variants of the CTMC method have been developed and employed to investigate ionisation collisions between many-electrons systems. The proposed approximations include the simple neglection of the target electron's interactions in the independent electron model, more elaborated models that include dynamical screening, energy- or momentum-dependent potentials that incorporate the quantum-mechanical uncertainty relations, and  semiclassical time-propagations \cite{Schultz1990JPBp3839,Schultz1992JPBp4601,Cohen1996PRAp573,Wood1999PRAp1317,Wood1997PRAp3746,Fiol2001JPBpL503,Geyer2003JPBpL107,Abbas2008PMBp41}. 
In particular $n$-CTMC, where interaction between electrons belonging to the same center are neglected but interactions between electrons of different atoms are fully included, provides a simple framework to investigate the relative importance of nucleus-electron to electron-electron interactions in collisions between dressed ions and atoms \cite{Fiol2001JPBpL503,Fiol2003JPBpL99}.

The quantum distorted wave models presented in~\cite{Monti2008JPBp201001,Monti2011JPBp195206} were extensions of the well-known Continuum Distorted Wave (CDW) \cite{Belkic1978JPBp3529,Crother1983JPBp3229} and Continuum Distorted Wave-Eikonal Initial State (CDW-EIS) \cite{Fainste1988JPBp287} models to the case of dressed projectiles.
Such extensions were obtained by approximating the interaction of the projectile with the active target-electron by an analytic two-parameter Green-Sellin-Zachor (GSZ) model potential \cite{Green1969PRp184,Szydlik1974PRAp1885,Garvey1975PRAp1144}. In these models it was also assumed that the projectile electrons are bound deeply enough to be considered as passive electrons (not changing their state during the collision).  They were successfully applied to study several systems, among which we can mention the ionisation of He atoms by $1$~MeV/u U$^{21+}$ and $600$~keV/u Au$^{11+}$ impact (see \cite{Monti2008JPBp201001} and \cite{Monti2011JPBp195206}, respectively).  In all the collision systems reported in \cite{Monti2008JPBp201001,Monti2011JPBp195206} the projectile internal structure was considered ``frozen'' and projectile ionisation was neglected.

In the present case the projectile electrons have binding energies similar to the binding energies in the helium target.  Therefore target, projectile and even simultaneous ionisation may give a substantial contribution to the electron emission process.

We present in this work a new set of experimental absolute DDCS for electron emission in collisions of $440$~keV/u Li$^{q+}$ ($q=1,2$) ions impinging over atomic He targets, as a function of electron energy for a wide range of fixed emission angles.  We compare the experimental data with the above mentioned four-body CTMC, CDW and CDW-EIS theoretical models.  Target, projectile and simultaneous ionisation are theoretically calculated and their relative contribution to electron emission is analysed. In the present analysis, target ionisation simultaneous with electron capture by the projectile, resulting on $\mathrm{He}^{2+}$ residual ions, has been neglected since its probability is small at the energies considered. In fact, total cross sections for this process account for less than 0.04\% of the total electron emission \cite{Woitke1998PRAp2692}.

Atomic units (a.u.) are employed except where otherwise stated.

%%%%%%%%%%%%%%%%%%%%%%%%%%%%%%%%%%%%%%%%%%%%%%%%%%%%%%%%%%%%%%%%%%%%%%

\section{Experimental arrangement and procedures}
\label{S:exp_setup}

Experimental data were measured with the 1.7~MV Tandem accelerator at Centro At\'{o}mico Bariloche. The experimental setup is described in detail elsewhere \cite{Bernard1996RSIp1761,Monti2012JPBp5202}, so only details related to the present experiment will be given here. Lithium ion beams emerging from the accelerator with 440 keV/u are collimated in the transport section by two sets of four collimators to $0.60~\times~0.60~\mathrm{mm}^2$, which determine a beam divergence of 0.7 mrad (half-angle). Inside the collision chamber the projectiles collides with the effusive target at the focus of a cylindrical mirror spectrometer. The analyzer rotates in a plane perpendicular to the gas flow direction so that it measures energy distributions of the electrons emitted in the collision for any selected angle between 0 and 180 degrees. After the collision the beam is collected in a Faraday cup. The collected charge is used to normalized the electronic distributions to a constant number of projectiles.

The present data were taken under the same conditions as the ones described in \cite{Monti2012JPBp5202}.The background pressure in the collision chamber was below $5 \times10^{-7}$~mbar, while the pressure in the transport section was lower than $4 \times10^{-6}$~mbar. The pressure in the collision chamber with target gas was set to $4.4 \times 10^{-5}$~mbar. The angular acceptance and energy resolution of the spectrometer were selected to be 2 degrees and 6\%{}, respectively.

Auxiliar data with uniform gas target density in the collision chamber were taken to correct  the distortions in the measured spectra due to the distribution of gas established in the chamber from the (localized) effusive  target. Relative normalization between different emission angles was estimated to have an uncertainty of $\pm 15\%$. Statistical errors are relatively important at low counting rate for the high energy range and at backward emission angles.

Doubly differential cross sections (DDCS) for electron emission in collisions of 440 keV/u Li$^+$ and Li$^{2+}$ with Helium atoms were determined from the electronic distributions \cite{Bernard1996RSIp1761}. They were measure ad angles, relative to the incident beam direction, $\theta=0$, 10, 20, 30, 40, 50, 70, 90, 120, 150, and 170 degrees.

%revisar
Normalization to absolute values was done by using total cross sections measured by Shah and Gilbody \cite{Shah1985JPBp899} and Woitke \emph{et~al.}~\cite{Woitke1998PRAp2692} for Li$^{3+}$+He. The discrepancies observed between these two sets of experimental data prevent us of using
directly the data taken by Woitke \emph{et~al.} for Li$^+$ and Li$^{2+}$.

In order to determine absolute cross section values, energy distributions for one selected angle were taken for Li$^{q+}$ ($q=1,2,3$) incident on He with the same energy and under the same experimental conditions. Numerical estimations of the beam charge fractions entering at the Faraday cup showed that
% (0.9996 for Li+ and 0.9998 for Li2+),
the contributions of the beam contaminants were less than 0.04\% for Li$^+$ and 0.02\% for Li$^{2+}$ beams. Therefore, as the charge collected in the FC is determined only by the primary beam, the number of projectiles was obtained by dividing the collected charge by the charge state $q$. Then the distributions were normalized to the same number of proyectiles. Using DDCS values already determined for Li$^{3+}$ \cite{Monti2012JPBp5202}, absolute DDCS were calculated for the other projectiles. An uncertainty of $\sim20\%$ was assessed from the disagreement between the available total cross section (TCS) data.

%%%%%%%%%%%%%%%%%%%%%%%%%%%%%%%%%%%%%%%%%%%%%%%%%%%%%%%%%%%%%%%%%%%%%%
\section{Theory}

In many-body collision systems with \textit{dressed} projectiles, such as those investigated in this work, ionisation may take place due to different identifiable physical mechanisms.  The electrons play a dual role: on one hand they screen the nucleus charge while on the other hand the direct electron-electron interaction influences the dynamics and may produce ionisation.
At high impact energies, the role of the interaction between target electrons and projectile electrons may be separated from the role of the electron-nucleus interactions, and they have been interpreted as two different mechanisms termed \textit{screening} and \textit{antiscreening} \cite{Mcguire1981PRAp97,Montene1994PRAp3186}.
In this context, antiscreening is associated with the collision of a quasi-free electron accompanying one center with the bound electron in the other. Thus, such a clear distinction can only be accomplished in fast collisions, when the binding energy is negligible compared to the electron's kinetic energy.

The measured spectra for the systems studied in the present work, obtained by electron-spectroscopy, are the result of the combined contributions from target-ionisation (eTI), projectile-ionisation (ePT), and simultaneous ionisation from both centers (TPI). In order to describe these channel we have performed theoretical calculations using two different quantum models, the Continuum Distorted Wave (CDW) and the Continuum Dirstorted Wave - Eikonal Initial State, and a classical approach using Classical-Trajectory Monte-Carlo simulations.

%%%%%%%%%%%%%%%%%%%%%%%%%%%%%%%
\subsection{Continuum distorted wave models for dressed-ion--impact single ionisation of atoms}
\label{dwmodels}
The electron emission process was studied by means of extensions of CDW and CDW-EIS models to describe ionization by dressed-projectiles.  The extensions of these theoretical models have been used previously to compute DDCS for several systems, combining different projectiles colliding with He targets at intermediate-high impact energies (see \cite{Monti2008JPBp201001,Monti2011JPBp195206}).

In order to treat multiple-electron systems within the independent electron model, we consider only one active electron and, following the procedure given in \cite{Fainste1988JPBp287} (see also~\cite{Stolter1997_EEI,Fainste1991JPBp3091}), the multielectronic Hamiltonian is reduced to:
\begin{equation}
  H_{el} = -\frac{1}{2} \nabla^{2} + V_{T}(\bi{x}) + V_{P}(\bi{s}) + V_{s}(\bi{R})\,,
\end{equation}
where $\bi{x}$ and $\bi{s}$ are the positions of the active target-electron in the target and projectile reference frames, respectively.  $V_{T}(\bi{x})$ is a model potential  taking into account the interaction of this electron with the remaining --partially dressed-- target, $V_{P}$ is the interaction between the projectile and the active electron that, according to the work presented in \cite{Monti2008JPBp201001,Monti2011JPBp195206}, is approximated with an analytical two-parameter GSZ potential:
\begin{eqnarray}
  V_{P}(\bi{s})=-\frac{q}{s}-\frac{1}{s}(Z_p-q)\left[H(e^{s/d}-1)+1\right]^{-1}
\, ,
\end{eqnarray}
where $q$ is the net (asymptotic) charge of the projectile, $Z_{P}$ is its nuclear charge, and $H$ and $d$ are parameters that depend on $Z_{P}$ and $q$.  $V_{s}(\bi{R})$ is the mean interaction of the projectile with the target nucleus and the passive electrons.  This last potential depends only on the internuclear coordinate $\bi{R}$ and thus, within the straight-line version of the impact-parameter approximation, produces a phase factor which does not affect the electron dynamics.  For both quantum approximations the initial bound state of the target was described by means of Roothaan-Hartree-Fock (RHF) wavefunctions~\cite{Clement1974ADNDTp445}.

Within the distorted-wave framework, the transition matrix may be written in its \emph{prior} or \emph{post} version, depending on whether the perturbative operator acts on the initial- or the final-channel distorted wavefunction.
The \emph{prior} version of the CDW model presents well-known difficulties in its computation due to a logarithmic divergence near the binary encounter structure (see \cite{Brauner1992PRAp2519}).
Consequently, in this work we employed the \emph{post} version, where the initial and final distorted-waves are given by:
\begin{eqnarray}
  \chi_{i}^{+}(\bi{x},t) & = & \Phi_{i}(\bi{x},t)\, \mathcal{L}_{i}^{+}(\bi{s})
  \label{dwini} \\
  \chi_{f}^-(\bi{x},t) & = & \Phi_{f}(\bi{x},t)\, \mathcal{L}_{f}^-(\bi{s})
  \label{dwfin}\, .
\end{eqnarray}
Here $\Phi_{i}(\bi{x},t) = \phi_{i}(\bi{x}) \exp{(-\rmi\varepsilon_{i} t)}$ and $\Phi_{f}(\bi{x},t) = \phi_{f}(\bi{x})\exp{(-\rmi\varepsilon_{f} t})$ are the initial-bound and final-continuum states which are solutions of the time-dependent Schr\"{o}dinger equations:
\begin{equation}
  \label{schi}
  \left[
    -\frac{1}{2} \nabla^{2} + V_{T}(\bi{x}) -\rmi\frac{\partial}{\partial t}\bigg|_{\bi{x}}
  \right]
  \Phi_{i,f}(\bi{x},t)= \varepsilon_{i,f}\Phi_i{,f}(\bi{x},t)\, .
\end{equation}
Also, $\varepsilon_{i}$ and $\varepsilon_{f}$ are the electron energies in the initial and final states, respectively.  We consider a RHF initial wavefunction and an effective Coulomb target potential $V_{T}(\bi{x})=-Z_{eff}/x$ in the final state.  Therefore, using the Belki\'{c} \textit{et~al.}~prescription \cite{Belkic1979PRp279}, an effective charge $Z_{eff}=\sqrt{-2\,n^2\,\varepsilon_{i}}$ is chosen in the residual target final continuum state, where $n$ is the principal quantum number of the $\phi_{i}$ orbital.  This effective charge partially considers the dynamic screening in the exit channel (see \cite{Monti2010JAMOPp128473,Monti2010JPBp205203}).  The initial distortion is proposed as:
\begin{equation}
  \label{inidist_cdw}
  \mathcal{L}_{i}^{+}(\bi{s})=N(\nu)\, \,_{1}F_{1}(\rmi\nu;1;\rmi vs + \rmi \bi{v}\cdot\bi{s} )\,
\end{equation}
whereas the final distortion is chosen as:
\begin{equation}
  \label{findist}
  \mathcal{L}_{f}^-(\bi{s})  = N^*(\zeta) \,_{1}F_{1}(-\rmi\zeta;1;-\rmi ps - \rmi \bi{p}\cdot\bi{s} )
\end{equation}
where $\bi{v}$ is the projectile velocity, $\nu = Z_{P}/v$, $\zeta=Z_{P}/p$, $\bi{p} = \bi{k} - \bi{v}$ is the ejected electron momentum in the projectile reference frame, being $\bi{k}$ the ejected electron momentum in the target reference frame, and $_{1}F_{1}$ is the confluent hypergeometric function.

The CDW-EIS results were obtained using the \textit{prior} version.  The main difference with the CDW approximation is that in the CDW-EIS the initial distortion is proposed as:
\begin{equation}
  \label{inidist_cdweis}
  \mathcal{L}_{i}^{+}(\bi{s})=\exp\left(-\rmi\nu \ln(vs+\bi{v}\cdot\bi{s})\right)
\end{equation}
instead of that given in eq.~\eref{inidist_cdw}.

In order to calculate electron-loss from the projectile the reaction is reversed and a reference frame transformation is applied from the projectile to the laboratory frame.  The electron-loss DDCS as a function of the energy $\varepsilon$ and angle $\theta$ of the emitted electron can thus be written in the laboratory reference frame as (see for example appendix~B of~\cite{Stolter1997_EEI}):
\begin{equation}
  \label{ddcs_transform}
  \frac{d \sigma(\varepsilon,\theta)}{d \Omega d \varepsilon}=  \left(\frac{\varepsilon}{\varepsilon'}\right)^{1/2}
  \frac{d \sigma(\varepsilon',\theta')}{d \Omega' d \varepsilon'}
\end{equation}
where the primed (unprimed) quantities are associated with the projectile (laboratory) reference frame.
The transformation rules for energy and angle are easily derived:
\begin{equation}
  \label{ddcs_transform-energy}
  \varepsilon'=\varepsilon+T-2\left(\varepsilon T\right)^{1/2}\cos{\theta} \,,
\end{equation}
and
\begin{equation}
  \label{ddcs_transform-angle}
\theta'=\arccos\left(\frac{\varepsilon-T-\varepsilon'}{2\sqrt{T\varepsilon'}}\right)\,  ,
\end{equation}
where $T=v^2/2$.

To investigate the simultaneous ionisation (PTI) a four-body system with two active electrons should be considered, similarly to the CTMC calculations.  There exist previous works of distorted-wave models which deal with a four-body approximation of the ionisation process \cite{Belkic1997JPBp1731,Monti2009JPBp195201}, but in these cases both active electrons are initially bound to the target and, accordingly, they are affected by the same perturbative potential.  Differently, in the case of PTI one of the two active electrons is initially bound to the projectile and the other to the target, being thus both aggregates affected by different perturbative potentials, which complicates the computational solution of the problem.  Hence, in order to simplify the calculation of the contribution of PTI
a probabilistic approach was considered using one-active electron cross sections.  The DDCS for PTI was estimated as
\begin{equation}
  \label{ddcs_simult}
  DDCS^{\mathrm{(PTI)}}=
  \frac{DDCS^{(T)} TCS^{(P)}}{TCS^{(T)}+TCS^{(P)}}+
  \frac{DDCS^{(P)} TCS^{(T)}}{TCS^{(T)}+TCS^{(P)}}\,  .
\end{equation}
In eq.~(\ref{ddcs_simult}) $DDCS^{(T,P)}$ and $TCS^{(T,P)}$ stand for the DDCS and TCS for target ionisation and projectile electron-loss, respectively, at the given projectile velocity $v$.
This rough approximation may be valid when the binding energy of one of the collision centers is much higher than the other as in this case.
Despite the simplicity of this approach we have found that it gives a fair estimation of the contribution of the PTI to the total electron emission spectra \cite{Monti2013PSp14031}.
This is also confirmed with our present findings, where the PTI cross-sections compare fairly well with those obtained in a 4-body CTMC.

%%%%%%%%%%%%%%%%%%%%%%%%%%%%%%%

\subsection{CTMC}
\label{ctmc}
Additionally to the quantum calculations we have performed four-body Classical-Trajectory Monte-Carlo (CTMC) simulations of the $\mathrm{Li}^{q+} + \mathrm{He}$ ionisation collisions. The CTMC method is well-known and has been extensively documented \cite{Olson1977PRAp531,Stolter1997_EEI,Reinhol1986PRAp3859,Fiol2001JPBpL503}.
We model the collision problem as a four-body system consisting of two centers, each with one electron. Both the lithium-ion projectiles and helium targets are described as one-effective-electron systems, where the potential produced by both parent nuclei incorporate the screening due to ``passive'' electrons. The interactions are represented by means of two-parameters GSZ model potentials similar to those used in the quantum calculations.

The evolution of the system is solved numerically  for a large number of independent trajectories and the information is extracted by statistical methods \cite{Reinhol1986PRAp3859,Abrines1966PPSp861,Hardie1983JPBp1983,Fiol2000JPBp5343,Fiol2002JPBp1173}.
The total process consists of three distinctive stages: preparation of the system, evolution until convergence is achieved, and statistical analysis of the final state.

In the first step the system is prepared following a microcanonical ensemble. The initial position and velocity of the electron and the nucleus in the target are randomly chosen such that the energy is fixed to the helium binding energy $\varepsilon_{i}= -0.903$~a.u., whereas the momentum distribution resembles the quantum-mechanical momentum probability density of the atom \cite{Reinhol1986PRAp3859}. The helium target is modelled as an one-electron atom, where the core interacts with the active electron and the projectile through a non-coulomb central potential, as given in ref.~\cite{Garvey1975PRAp1144}.
The projectile initial velocity is fixed and determined by the collision energy, while its impact parameter is randomly chosen with a distribution that describes an uniform flux. The internal representation of the projectile $\mathrm{Li}^{q+}$ is prepared similarly than the target, using the experimental binding energy of the ion.

The classical evolution of the system is obtained by solving numerically Hamilton's canonical equations by means of a modified middle-point code with an adaptive step-size control.  When convergence is achieved within $0.02\%$, the velocities of the fragments in the final state are determined and the DDCS are evaluated by the formula 
\begin{equation*}
  \frac{\rmd \sigma}{\rmd \Omega \rmd E}= \frac{N_{i}(\Omega, E)/\Delta E\, \Delta \Omega}{N/(\pi\, b^{2}_{\mathrm{max}})}\,.
\end{equation*}
Here $N_{i}(\Omega, E)$ is the number of ionisation events where the electrons have energy in a neighborhood $\Delta E$ of $E$ and are emitted in a solid angle $\Delta \Omega$ in the direction of $\Omega$.  The number of ionisation events is normalized to the incident flux $N/(\pi\, b^{2}_{\mathrm{max}})$ where $b_{\mathrm{max}}$ is the maximum impact parameter evaluated, larger than the maximum impact parameter that produces ionisation.  In this work the number of trajectories solved is approximately $N=1.5 \times 10^{8}$ for each collision system.
The acceptances $\Delta \Omega$ in the solid angle and $\Delta E$ in the energy bins were chosen to coincide in each point with those employed in the experiment.

\section{Results and Discussions}

\subsection{Experimental data}

\begin{figure}
  \centering \includegraphics[width=.95\textwidth,clip]{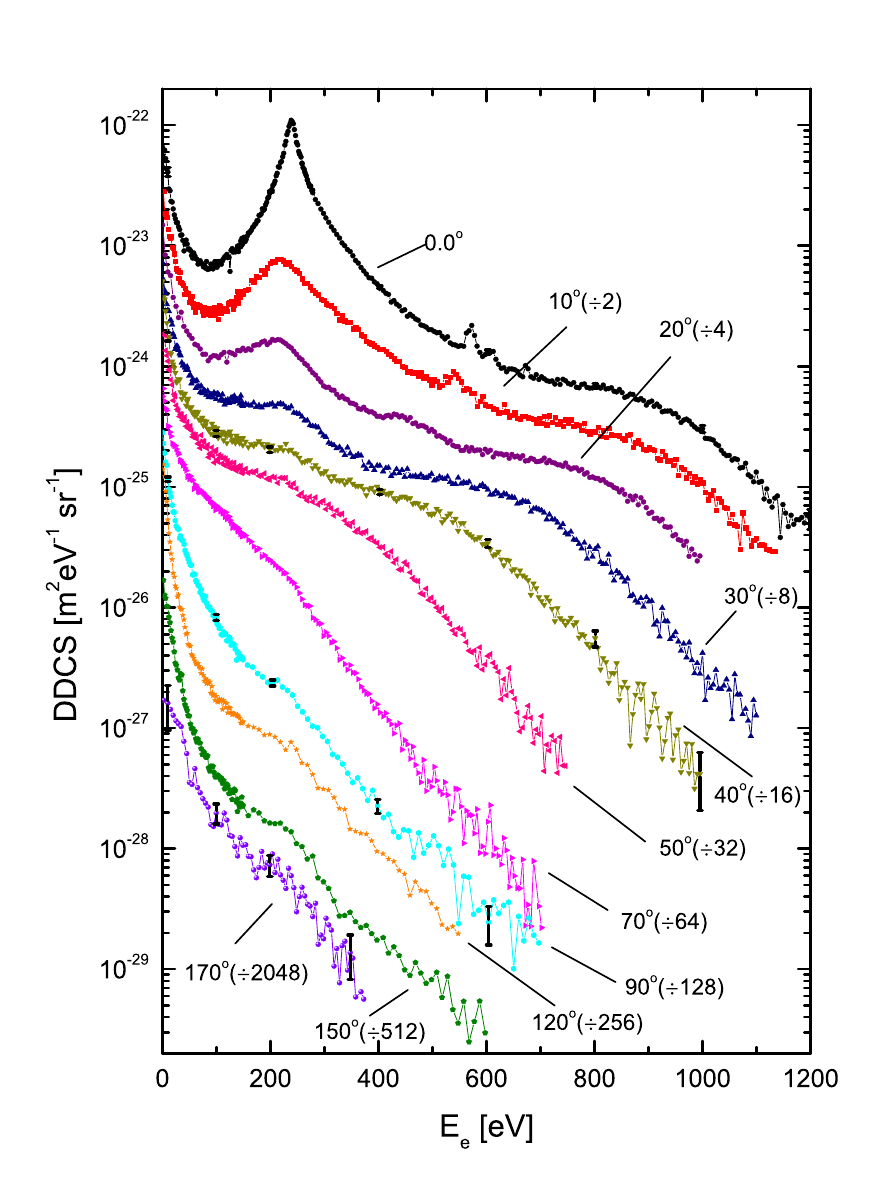}
\caption{Doubly differential cross-section for electron emission in
collisions of 440~keV/u Li$^{+}$ on He targets as a function of the electron energy.
Emission angles are indicated in the figure. Different factors for each curve, shown in the figure, were applied for a better display.
Typical error bars are displayed for $\theta=0^{\circ}$, 40$^{\circ}$, 90$^{\circ}$, 150$^{\circ}$ and 170$^{\circ}$.
\label{limheexp}
}
\end{figure}

\begin{figure}
  \centering \includegraphics[width=.99\linewidth,clip]{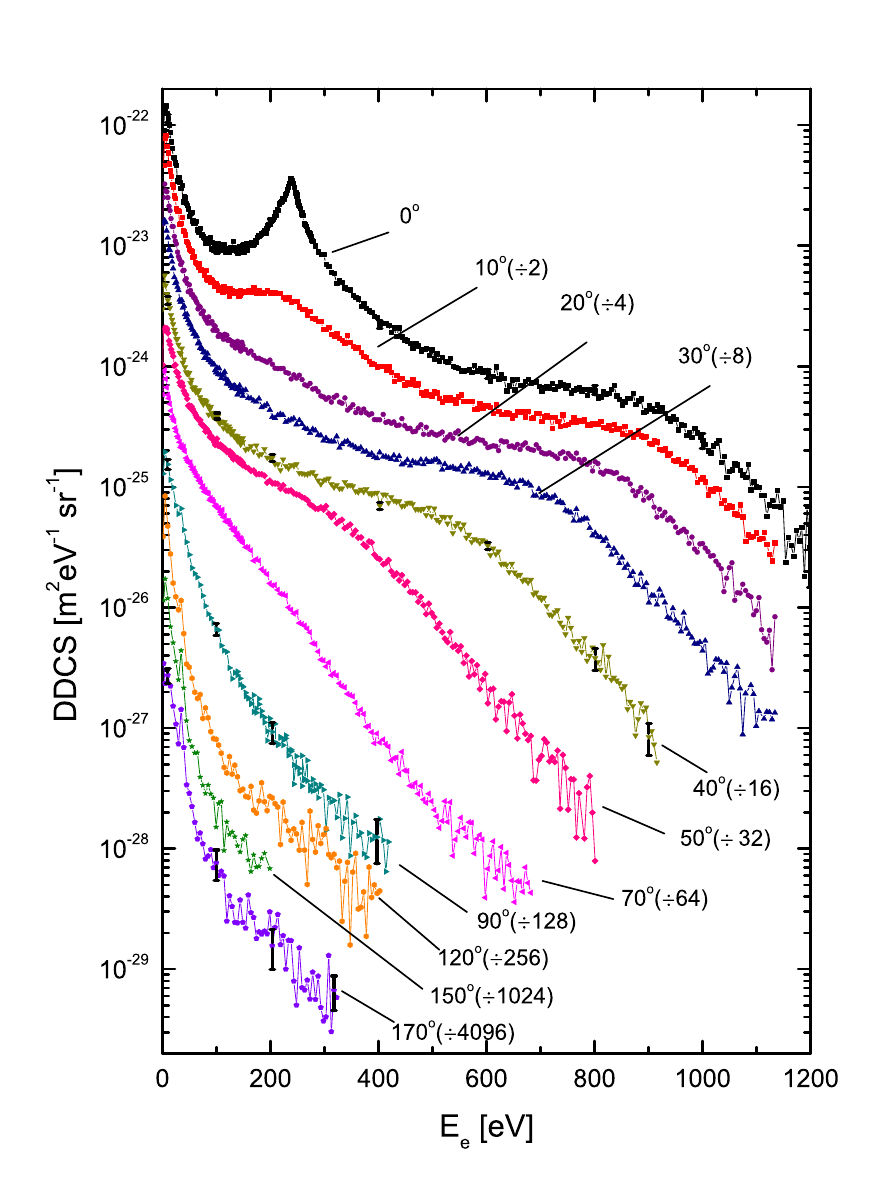}
\caption{
Same as in figure \ref{limheexp} for collisions of 440~keV/u Li$^{2+}$ with He targets.
\label{li2mheexp}
}
\end{figure}

Experimental absolute DDCS for electron emission in collisions of Li$^+$ and Li$^{2+}$ with helium targets are shown in figures \ref{limheexp} and \ref{li2mheexp}. For display purposes the data have been divided by different factors for different emission angles, as indicated in the figures. In the spectra for 0$^\circ$, 40$^\circ$, 90$^\circ$, 150$^\circ$ and 170$^\circ$, we have included some bars indicating the statistical errors for different emission-energy regions.

A comparison between the total cross section values for the different channels contributing to the total emission shows that the dominant process is the target ionization, which represents 73\% of the total emission for Li$^+$ and 95\% for Li$^{2+}$ \cite{Woitke1998PRAp2692}. The remaining electron emission is produced by projectile and simultaneous ionization, while transfer-ionization (electron capture simultaneous with target ionization) processes are negligible at this energy.

The main structure observed at $\theta = 0^\circ$ and centered at $E_{e} \approx 240$ eV is formed by the superposition of target-electrons captured to the continuum by the projectile (ECC) and electron-loss by the projectile to the continuum (ELC). As it is expected due to its lower ionization energy, the intensity of the electron-loss peak is much more important for incident $\mathrm{Li}^{+}$ than for $\mathrm{Li}^{2+}$. The yield of the structure decreases for higher angles and its maximum is located in a ring centered at zero energy with a radius, that for fast collisions is slightly smaller than the kinetic energy $E_{e}\approx v^{2}/2$, associated with the incident velocity $\bi{v}$. This ridge is formed by the emission of projectile-electrons after a binary-encounter interaction (PBE) with the target.

% Figure 1
%
% Figure 2

At $\theta=0^{\circ}$ and 800~eV another broad ridge structure is present that is associated with target-electrons emitted in binary-encounter collisions with the projectile (TBE). The position of this structure shifts to lower values when increasing the emission angle as it is expected~\cite{Stolter1997_EEI}.

At low emission angles a set of peaks at $E_e \approx 560$~eV is observed for ionization by Li$^+$. They are produced by autoionization emission from doubly-excited Li$^+$ projectiles~\cite{Bruch1975PRAp1808}.

In the case of $\mathrm{Li}^{2+}$ projectiles this emission could only occur after single electron capture, whose cross-section is small at the present collision velocity, being not visible in the experimental data.

\begin{figure}
  \centering \includegraphics[width=.99\textwidth,clip]{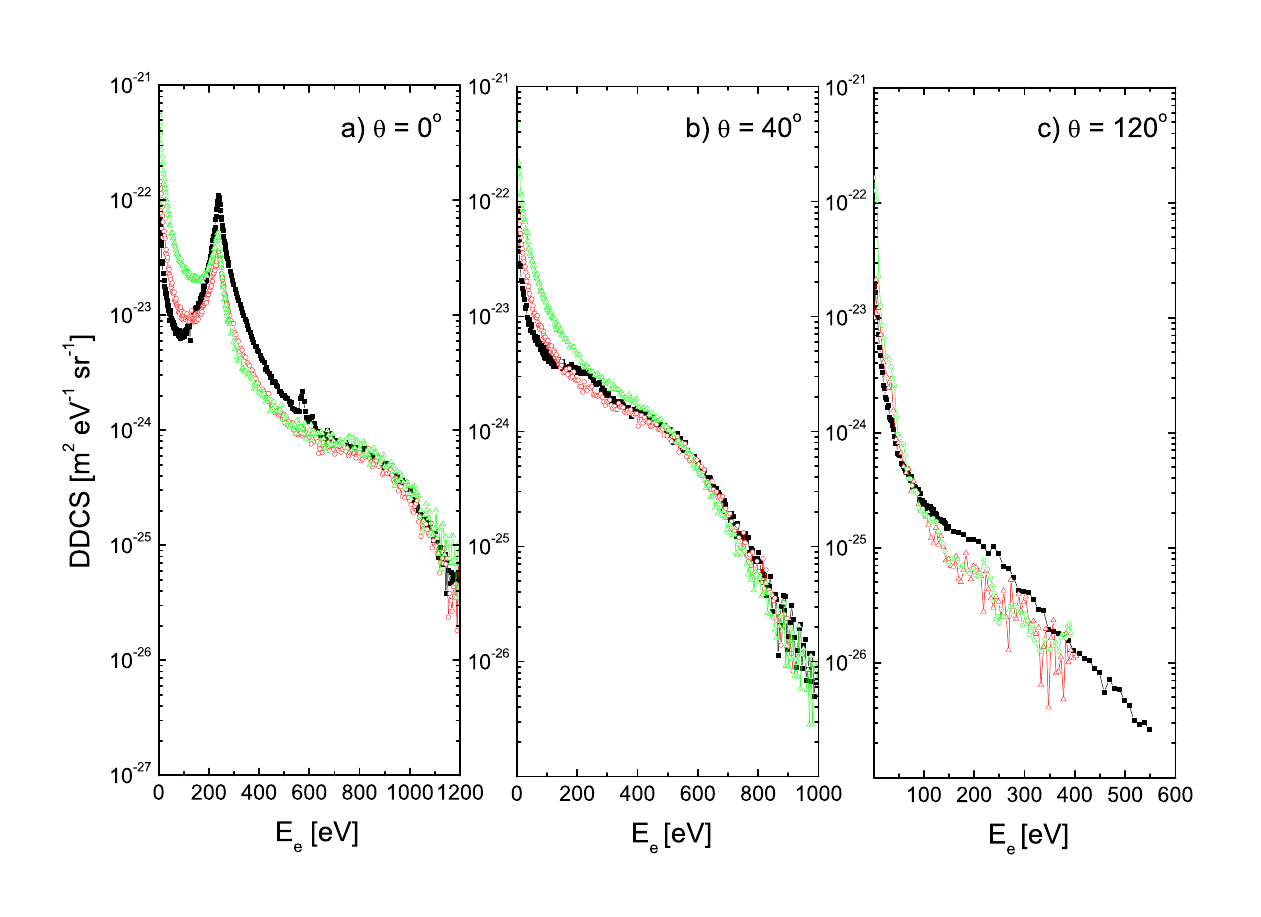}
\caption{Doubly differential cross-sections for electron emission in
collisions of 440~keV/u Li$^{q+}$ (q=1,2,3) on He targets as a function of the electron energy
for three different emission angles:
a) $\theta = 0^\circ$,
b) $\theta = 40^\circ$ and
c) $\theta = 120^\circ$.
\fullsquare: Li$^{+}$;
\textcolor{red}{\opencircle}, Li$^{2+}$ ;
\textcolor{green}{\opentriangle}, Li$^{3+}$ \cite{Monti2012JPBp5202}.
\label{liqmheexp}
}
\end{figure}

The effect of the projectile charge-state on the DDCS varies for different energies and angles of the emitted electrons.
In figure \ref{liqmheexp} electronic distributions for collisions of 440~keV/u Li$^{q+}$, $q$=1,2,3 with He atoms are shown for three angles of emission, $\theta = 0^\circ, 40^\circ$ and $120^\circ$. Fast electron emission ($E_e \gtrsim E_{BE}$) in the forward direction, shows no dependency on the initial projectile charge state, which is consistent with the characteristic low impact-parameter associated. In the backward direction however, the largest emission occurs for Li$^+$ projectiles, which is expected to come from the binary-encounter collisions between projectile-electrons and the target in the forward direction of the projectile frame.

% Figure 3

On the other hand, the emission of electrons with low energy is mainly due to the ionization of the target. It is strongly dependent on the initial projectile's charge-state, with larger emission probabilities for higher charged projectiles. Since the release of slow-electrons is dominated by long-distance interactions, for a given velocity the dominant parameter is the initial projectile's charge-state.

The main contribution of projectile electrons to the total emission occurs mainly at the electron-loss peak and the corresponding binary-encounter ridge (PBE). In the backward direction PBE is the dominant emission process at high energies.
However, in the case of Li$^{2+}$, the contributions of projectile electrons are very small and the overall emission spectra is similar to the one observed for Li$^{3+}$. This may be attributed to the large value of the ionisation energy of Li$^{2+}$, $\varepsilon= 122.4$~eV.

\subsection{Comparison with theoretical results}

Theoretical calculations and experimental results are compared in figures \ref{li1mheth} and \ref{li2mheth} for both ions, Li$^+$ and Li$^{2+}$, and for three typical emission angles, $\theta = 10^\circ, 52^\circ$ and $150^\circ$. In each case, the contribution from electron-target ionization (eTI), electron-projectile ionization (ePI) and simultaneous ionization from both centers (PTI) are shown.

\subsubsection{Li$^+$ projectiles}~

\begin{figure}
  \centering \includegraphics[width=.95\textwidth,clip]{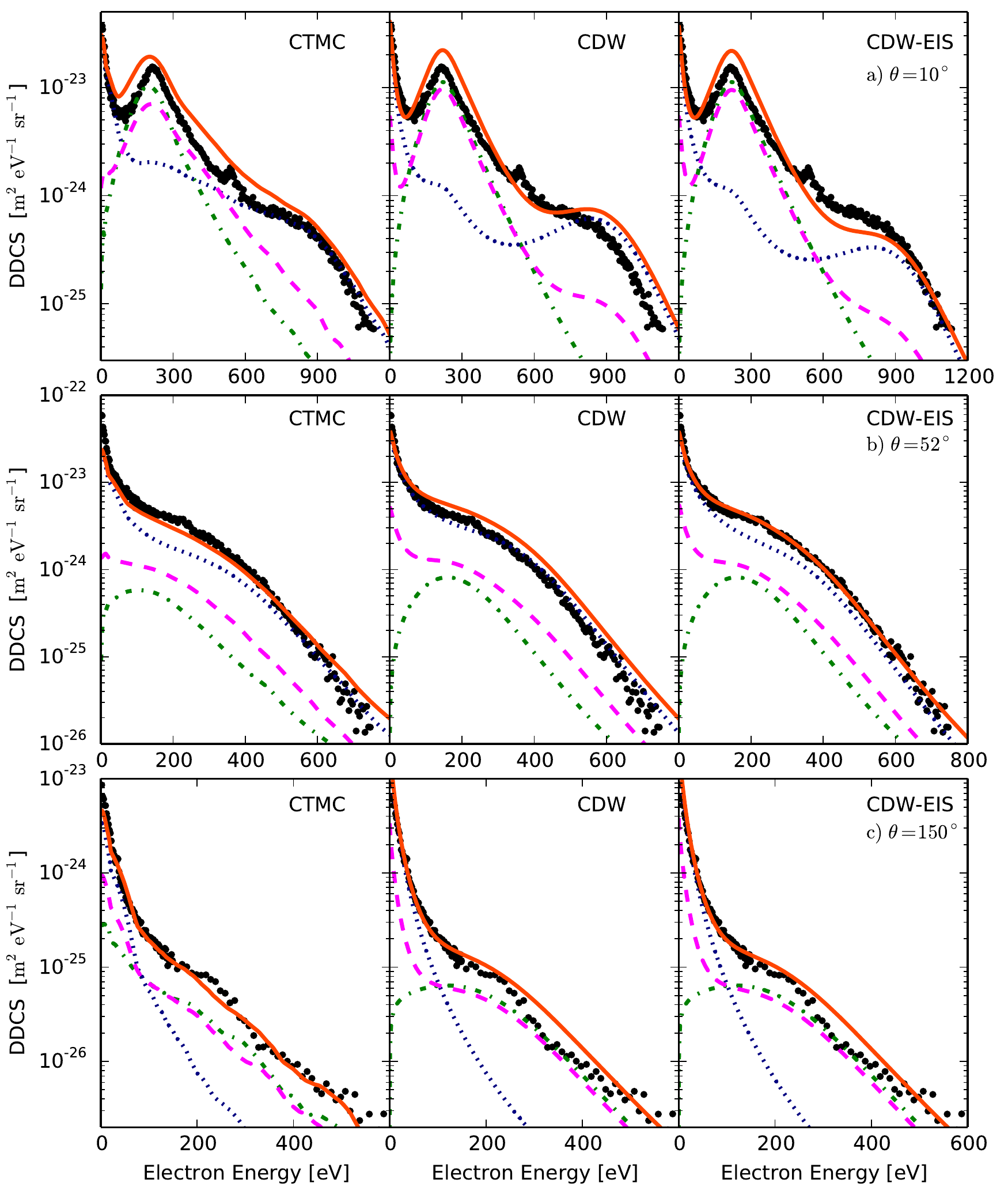}
\caption{Doubly differential cross-sections for electron emission in
collisions of 440~keV/u Li$^{+}$ on He targets as a function of the electron energy for three different emission angles:
a) $\theta = 10^\circ$,
b) $\theta = 52^\circ$ and
c) $\theta = 150^\circ$.
\fullcircle: Experiment;
\textcolor{blue}{\dotted}, eTI;
\textcolor{green}{\chain}, ePI;
\textcolor{magenta}{\dashed}, PTI;
\textcolor{orange}{\full}, Total emission.
\label{li1mheth}}
\end{figure}

% The present calculations for simultaneous ionization are much larger than experimental total cross-sections estimated from the work by Woitke \etal \cite{Woitke1998PRAp2692} (see tables \ref{tcslim} and \ref{tcsli2m}).

The three theoretical approximations reproduce fairly well the experimental data.
They show similar qualitative behaviour as a function of the electron energy, but CDW-EIS calculations give a slightly better agreement with the experiment in the case of Li$^+$ (\fref{li1mheth}) than the other two theories.
In all cases the larger differences appear at small emission angles. Besides a slight overall overestimation of CTMC, the CDW and CDW-EIS approximations present a marked shoulder due to target binary-encounter (TBE) processes, that appears perceptibly milder in the data. 
At intermediate and large angles the agreement of the theories with the experiment improves noticeably. The observed discrepancies may arise from the approximated theoretical treatment of the target and projectile as one-electron systems, and the corresponding complete neglection of correlations between electrons in the same center. Additionaly, the quantum theories do not include the interactions between electrons belonging to different centers.

The contributions of the different channels included in the calculations to the total emission depend strongly on the region of energy and angles investigated. For high-energy electrons, eTI is the major contributor for emission in the forward direction, while ePI and PTI are dominant for backward emission.
For electrons emitted with velocities close to the incident velocity the main contribution in the forward direction comes from projectile electrons (ELC), ionized either by interaction with the core target or by electron-electron interaction.  The three theories show that ePI contributes minimally, accounting for approximately only $10\%$ of the total emission up to electron angles of 20$^{\circ}$.
This result is consistent with our previous analysis: besides ELC, electrons from the projectile are emitted in the PBE ridge, and its contribution is important only in the backward direction, where the target ionisation is very small. Thus, ePI and PTI dominates the backward, high-energy emission region.

The electron energy range where eTI contribution is the most important consistently shrinks for increasing emission angles and, at 90$^\circ$, the electron emission for $E_e \gtrsim$ 200 eV is dominated by ePI and PTI. In backward direction, as the electron energy considered increases, all the theories predict dominant contributions from PTI and ePI.

\subsubsection{Li$^{2+}$ projectiles} 
~

\begin{figure}
  \centering \includegraphics[width=.96\textwidth,clip]{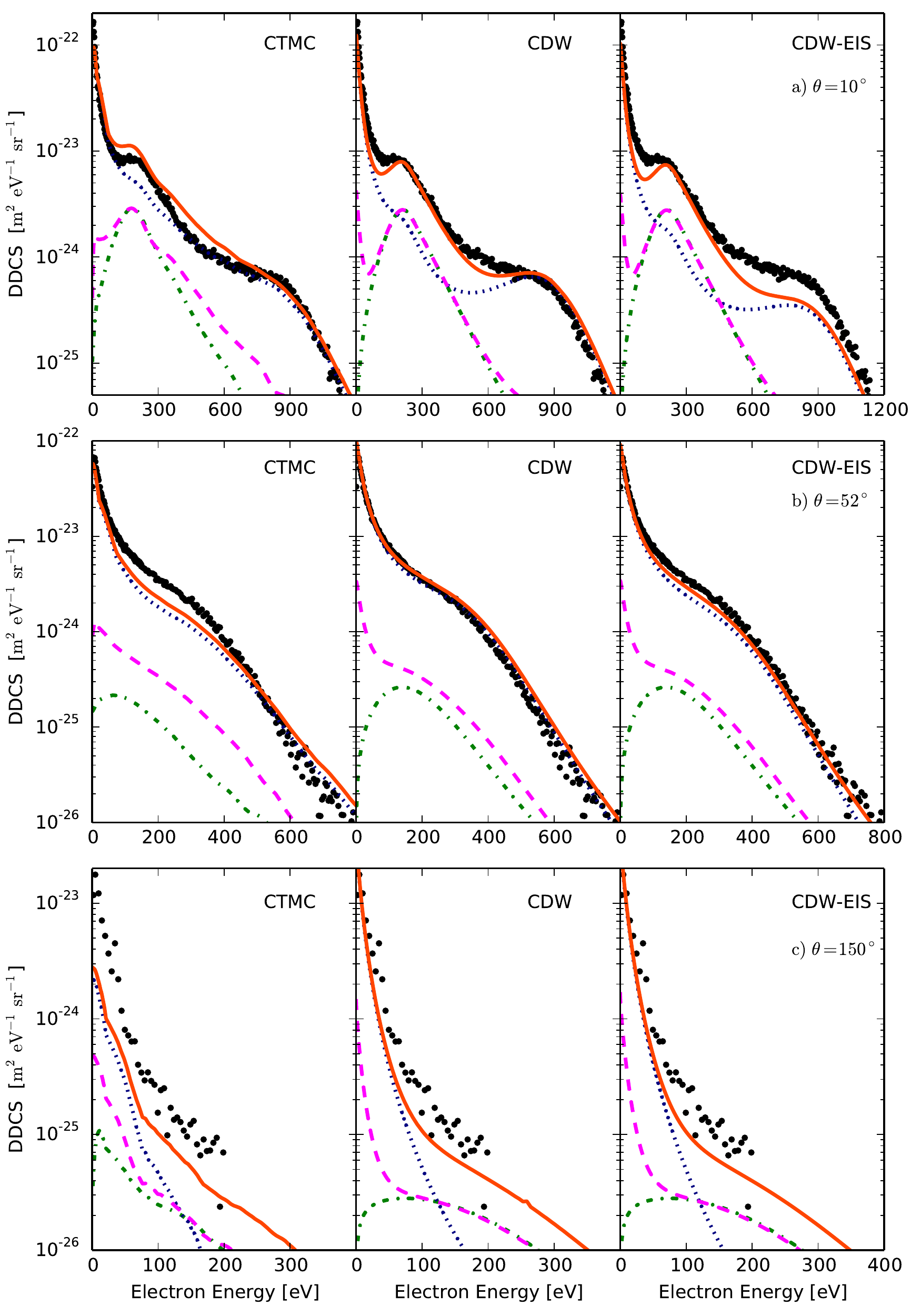}
\caption{Doubly differential cross-section for electron emission in
collisions of 440~keV/u Li$^{2+}$ on He targets as a function of the electron energy for three different emission angles:
a) $\theta = 10^\circ$,
b) $\theta = 52^\circ$ and
c) $\theta = 150^\circ$.
% Left panel, CTMC results; Center panel, CDW results, and right panel, CDW-EIS calculations.
\fullcircle: Experiment;
\textcolor{blue}{\dotted}; eTI;
\textcolor{green}{\chain}: ePI;
\textcolor{magenta}{\dashed}: PTI;
\textcolor{orange}{\full}: Total emission.
\label{li2mheth}}
\end{figure}

\Fref{li2mheth} shows the comparison of experimental cross-section data with calculations for Li$^{2+}$ projectiles. The conditions are the same than for $\mathrm{Li}^{+}$ in \fref{li1mheth}.
As expected, due to the higher ionization energy, ePI and PTI are relevant only in the region of EL peak, but they become dominant for fast electrons in the backward direction.

As in the $\mathrm{Li}^{+}$ case, the overall agreement between calculations and experiment, as well as between different theoretical  approaches, is remarkably good. CTMC somewhat overestimates the DDCS at small angles, showing a binary-encounter (TBE) structure that is less pronounced than the showed in the experimental data. On the contrary, CDW calculations present a more marked TBE structure than the data.

As in the case of $\mathrm{Li}^{+}$ the remarkable agreement between the three theories, is observed not only on the total emission cross sections but also on each of the processes contributing to the measured spectra.
Both quantum theories show a very similar behavior of the doubly-differential cross-sections for target, projectile and simultaneous ionization. CTMC dependence of DDCS are also very similar to the quantum results but simultaneous ionization is slightly wider at small angles, extending to higher emission energies.

It must be noted that in both quantum theories the electron-electron interaction is included in the nucleus-electron model potentials, as a screening of the nuclei charge. On the other hand, in the CTMC calculations the interaction between active electrons is included separately from the screened nucleus interactions. However,
the similarity of the spectra in the different theories suggests that the main effect of the electron-electron interaction is through the screening of the nucleus. Separation of simultaneous ionization spectra in contributions  by target and projectile ionization confirms this result. Though they are not shown, the contribution to PTI of electrons emitted from the target resembles single target ionization (eTI) spectra while the contribution from the projectile-electrons is similar to ePI.

\subsection{Total Cross Sections}

%As it was mentioned above in section \ref{S:exp_setup}, measured distributions were normalized to absolute DDCS by comparison with data for total ionisation cross-sections for stripped $\mathrm{Li}^{3+}$ ions taken by Woitke \etal \cite{Woitke1998PRAp2692}.

Results for total cross sections of the different channels contributing to the total electron emission are shown in tables \ref{tcslim} and \ref{tcsli2m}.
\begin{table}[!Hbpt]
\begin{center}
\caption{\label{tcslim} Total cross sections (in units of $10^{-20}\mathrm{m}^{2}$) for electron emission in collisions of 440 keV/u Li$^+$ with He. Data labeled as Woitke \etal are estimations from ref.~\cite{Woitke1998PRAp2692}}
\begin{tabular}{|c|c|c|c|c|c|}
 \hline
 Process & Woitke \emph{et al.} & This exp. & CTMC & CDW & CDW-EIS \\
 \hline
Total emission & $1.03 \pm 0.3 $  & $1.4 \pm 0.3$ & $1.57$  & $1.82$ & $1.65$  \\
eTI            & $0.70 \pm 0.2$  & $-   $ & $0.82$  & $1.23$ & $1.07$  \\
ePI            & $0.13 \pm 0.04$  & $-   $ & $0.18$  & $0.22$ & $0.22$  \\
PTI            & $0.065 \pm 0.019$  & $-   $ & $0.29$  & $0.37$ & $0.36$  \\
 \hline
\end{tabular}
\end{center}
% \end{table}
% %Li2+
% \begin{table}[!Hbtp]
\begin{center}
\caption{\label{tcsli2m} Total cross-sections (in units of $10^{-20}\mathrm{m}^{2}$) for electron emission in collisions of 440~keV/u Li$^{2+}$ with He. Data labeled as Woitke \etal are estimations from
\cite{Woitke1998PRAp2692}}
\begin{tabular}{|c|c|c|c|c|c|}
 \hline
 Process & Woitke et al. & This exp. & CTMC & CDW & CDW-EIS \\
 \hline
Total emission & $2.07 \pm 0.57$ & $1.90 \pm 0.4$ & $1.71$ & $2.13$ & $1.92$ \\
eTI            & $1.8 \pm 0.5$ & $-   $ & $1.36$ & $1.93$ & $1.72$ \\
ePI            & $0.033 \pm 0.012$ & $-   $ & $0.06$ & $0.07$ & $0.07$ \\
PTI            & $0.024 \pm 0.007$ & $-   $ & $0.14$ & $0.13$ & $0.13$ \\
 \hline
\end{tabular}
\end{center}
\end{table}

TCS values displayed in the first column are estimations from Woitke \emph{et al.} data \cite{Woitke1998PRAp2692} for 440~keV/u incident energy. Values obtained in the present experiment are shown in the second column.  Errors were mainly due to the normalization process, since other sources of error, like the integration method, are negligible by comparison. The other columns correspond to theoretical TCS values.
As the theories include only one active electron per collision center, tabulated experimental values for eTI, ePI and PTI from \cite{Woitke1998PRAp2692} do not include double electron emission from the same collision center.

TCS values for electron emission from the present experiment show good agreement with those obtained by Woitke \etal. In the case of the theoretical calculations, the agreement is good for Li$^{2+}$ projectiles while there is an overestimation for Li$^+$ incident on He.

The comparison between theoretical results is good for ePI and PTI channels, showing the main difference for eTI. There is a fair agreement between theories and tabulated values, with the higher differences appearing for the PTI cross-sections, where factors of about 5 or 6 are observed. Also, the quantum calculations are systematically higher than the classical results.

\section{Conclusions}
We have investigated experimentally and theoretically electron emission in collisions between helium and dressed projectiles of lithium. Experimental data on double differential cross-sections have been compared to a classical and two quantum-mechanical theories. The theoretical results are consistent with the present experimental data as well as with available published total cross-sections.
The reasonable accord between experimental and theoretical DDCS for total electron emission, as well as the general agreement between different theories, give us confidence to draw conclussions on the relative importance of the different channels involved in the ionization collisions. The present findings allowed us to determine the contribution of single target ionization, single projectile ionization, and simultaneous target-projectile ionization in different regions of the measured spectra.

\ack
JMM and RDR acknowledge the Agencia Nacional de Promoci\'{o}n Cient\'{\i}fica y Tecnol\'{o}gica for financial support through the project PICT2011-2045.
JF and PDF acknowledge support by the Consejo Nacional de Investigaciones Cient\'{\i}ficas y T\'{e}cnicas (Grant PIP 112-200901-00166).

%%%%%%%%%%%%%%%%%%%%%%%%%%%%%%%%%%%%%%%%%%%%%%%%%%%%%%%%%%%%%%%%%%%%%%%%%%
%%%% References
\bibliography{Paper_Liq+}

\begin{thebibliography}{10}

\bibitem{Stolter1997_EEI}
N.~Stolterfoth, R.~D. DuBois, and R.~D. Rivarola.
\newblock {\em Electron {E}mission in {H}eavy {I}on-{A}tom {C}ollisions}.
\newblock Springer-Verlag, Berlin, 1997.

\bibitem{Monti2012JPBp5202}
J.~M. Monti, D.~Fregenal, S.~Su\'{a}rez, P.~D. Fainstein, R.~D. Rivarola,
  G.~Bernardi, and J.~Fiol.
\newblock Experimental and theoretical results on electron emission from helium
  by the impact of bare {L}i$^{3+}$ ions.
\newblock {\em Journal of Physics B: Atomic Molecular and Optical Physics},
  45:5202, 2012.

\bibitem{Monti2008JPBp201001}
J.~M. Monti, R.~D. Rivarola, and P.~D. Fainstein.
\newblock F{A}{S}{T} {T}{R}{A}{C}{K} {C}{O}{M}{M}{U}{N}{I}{C}{A}{T}{I}{O}{N}:
  {Q}uantum interferences in swift highly-charged dressed-ion atom collisions.
\newblock {\em Journal of Physics B: Atomic Molecular and Optical Physics},
  41:201001, 2008.

\bibitem{Monti2011JPBp195206}
J.~M. Monti, R.~D. Rivarola, and P.~D. Fainstein.
\newblock Distorted wave theories for dressed-ion-atom collisions with
  {G}{S}{Z} projectile potentials.
\newblock {\em Journal of Physics B: Atomic Molecular and Optical Physics},
  44:195206, 2011.

\bibitem{Schultz1990JPBp3839}
D.~R. Schultz, R.~E. Olson, C.~O. Reinhold, S.~Kelbch, C.~Kelbch,
  H.~Schmidt-Bocking, and J.~Ullrich.
\newblock Coincident charge state production in {F}$^{6+}$+{N}e collisions.
\newblock {\em Journal of Physics B: Atomic Molecular and Optical Physics},
  23:3839--3847, 1990.

\bibitem{Schultz1992JPBp4601}
D.~R. Schultz, L.~Meng, and R.~E. Olson.
\newblock Classical description and calculation of ionization in collisions of
  100 e{V} electrons and positrons with {H}e and {H}2.
\newblock {\em Journal of Physics B: Atomic Molecular and Optical Physics},
  25:4601--4618, 1992.

\bibitem{Cohen1996PRAp573}
James~S. Cohen.
\newblock Quasiclassical-trajectory {M}onte {C}arlo methods for collisions with
  two-electron atoms.
\newblock {\em Physical Review A}, 54:573--586, 1996.

\bibitem{Wood1999PRAp1317}
C.~J. Wood and R.~E. Olson.
\newblock Double electron removal and fragmentation model of the {H}$_{2}$
  molecule by highly charged ions.
\newblock {\em Physical Review A}, 59(2):1317--1328, 1999.

\bibitem{Wood1997PRAp3746}
C.~J. Wood, R.~E. Olson, W.~Schmitt, R.~Moshammer, and J.~Ullrich.
\newblock Momentum spectra for single and double electron ionization of {H}e in
  relativistic collisions.
\newblock {\em Physical Review A}, 56:3746--3752, 1997.

\bibitem{Fiol2001JPBpL503}
J.~Fiol, R.~E. Olson, A.~C.~F. Santos, G.~M. Sigaud, and E.~C. Montenegro.
\newblock Simultaneous {P}rojectile and {T}arget {I}onization in {H}e$^{+}$ +
  {N}e {C}ollisions.
\newblock {\em Journal of Physics B: Atomic Molecular and Optical Physics},
  34(16):L503--L509, 2001.

\bibitem{Geyer2003JPBpL107}
T.~Geyer and J.~M. Rost.
\newblock Dynamical stabilization of classical multi-electron targets against
  autoionization.
\newblock {\em Journal of Physics B: Atomic Molecular and Optical Physics},
  36(4):L107--L112, 2003.

\bibitem{Abbas2008PMBp41}
I.~Abbas, C.~Champion, B.~Zarour, B.~Lasri, and J.~Hanssen.
\newblock N{O}{T}{E}: {S}ingle and multiple cross sections for ionizing
  processes of biological molecules by protons and $\alpha$-particle impact: a
  classical {M}onte {C}arlo approach.
\newblock {\em Physics in Medicine and Biology}, 53:41, 2008.

\bibitem{Fiol2003JPBpL99}
J.~Fiol, R.~E. Olson, R.~Moshammer, and J.~Ullrich.
\newblock Dynamical electron-electron correlation in {C}$^{2+}+${H}e
  simultaneous target-projectile collisional ionization.
\newblock {\em Journal of Physics B: Atomic Molecular and Optical Physics},
  36:L99--L105, 2003.

\bibitem{Belkic1978JPBp3529}
D{\v{z}}~Belki{\'c}.
\newblock {\em Journal of Physics B: Atomic Molecular and Optical Physics},
  11:3529, 1978.

\bibitem{Crother1983JPBp3229}
D.~S.~F. Crothers and J.~F. McCann.
\newblock Ionisation of atoms by ion impact.
\newblock {\em Journal of Physics B: Atomic Molecular and Optical Physics},
  16:3229--3242, 1983.

\bibitem{Fainste1988JPBp287}
P.~D. Fainstein, V.~H. Ponce, and R.~D. Rivarola.
\newblock A theoretical model for ionisation in ion-atom collisions.
  {A}pplication for the impact of multicharged projectiles on helium.
\newblock {\em Journal of Physics B: Atomic Molecular and Optical Physics},
  21(2):287--299, 1988.

\bibitem{Green1969PRp184}
A.~E.~S. Green, D.~L. Sellin, and A.~S. Zachor.
\newblock {\em Phys. Rev.}, 1:184, 1969.

\bibitem{Szydlik1974PRAp1885}
P.~P. Szydlik and A.~E. Green.
\newblock Independent-particle-model potentials for ions and neutral atoms with
  ${Z}\le 18$.
\newblock {\em Physical Review A}, 9:1885--1894, 1974.

\bibitem{Garvey1975PRAp1144}
R.~H. Garvey, C.~H. Jackman, and A.~E.~S. Green.
\newblock Independent-particle-model potentials for atoms and ions with
  $36<{Z}\le 54$ and a modified {T}homas-{F}ermi atomic energy formula.
\newblock {\em Physical Review A}, 12:1144--1152, 1975.

\bibitem{Woitke1998PRAp2692}
O.~Woitke, P.~A. Z\'{a}vodszky, S.~M. Ferguson, J.~H. Houck, and J.~A. Tanis.
\newblock Target ionization and projectile charge changing in 0.5-8-{M}e{V}/q
  {L}iq++{H}e (q=1,2,3) collisions.
\newblock {\em Physical Review A}, 57:2692--2700, 1998.

\bibitem{Bernard1996RSIp1761}
G.~Bernardi, S.~Su\'{a}rez, D.~Fregenal, P.~Focke, and W.~Meckbach.
\newblock Measurement of doubly differential electron distributions induced by
  atomic collisions: {A}pparatus and related instrumental effects.
\newblock {\em Review of Scientific Instruments}, 67:1761--1768, 1996.

\bibitem{Shah1985JPBp899}
M.~B. Shah and H.~B. Gilbody.
\newblock Single and double ionisation of helium by {H}+, {H}e2+ and {L}i3+
  ions.
\newblock {\em Journal of Physics B: Atomic Molecular and Optical Physics},
  18:899--913, 1985.

\bibitem{Mcguire1981PRAp97}
J.~H. McGuire, N.~Stolterfoht, and P.~R. Simony.
\newblock Screening and antiscreening by projectile electrons in high-velocity
  atomic collisions.
\newblock {\em Physical Review A}, 24:97--102, 1981.

\bibitem{Montene1994PRAp3186}
E.~C. Montenegro and T.~J.~M. Zouros.
\newblock Relationship between the {B}orn and impulse approximations for the
  antiscreening process.
\newblock {\em Physical Review A}, 50:3186--3191, 1994.

\bibitem{Fainste1991JPBp3091}
P.~D. Fainstein, V.~H. Ponce, and R.~D. Rivarola.
\newblock Two-centre effects in ionization by ion impact.
\newblock {\em Journal of Physics B: Atomic Molecular and Optical Physics},
  24(14):3091--3119, 1991.

\bibitem{Clement1974ADNDTp445}
C.~Clementi and C.~Roetti.
\newblock {\em Atomic Data Nuclear Data Tables}, 14:445, 1974.

\bibitem{Brauner1992PRAp2519}
M.~Brauner and J.~H. Macek.
\newblock Ion-impact ionization of {H}e targets.
\newblock {\em Physical Review A}, 46:2519--2531, 1992.

\bibitem{Belkic1979PRp279}
D{\v{z}}~Belki{\'c}, R.~Gayet, and A.~Salin.
\newblock {\em Physics Reports}, 56:279, 1979.

\bibitem{Monti2010JAMOPp128473}
J.~M. Monti, O.~A. Foj\'{o}n, J.~Hanssen, and R.~D. Rivarola.
\newblock A {C}omplete {P}ostversion of the {T}hree-{B}ody {C}ontinuum
  {D}istorted {W}ave-{E}ikonal {I}nitial {S}tate {A}pproximation for {S}ingle
  {I}onization of {M}ultielectron {A}toms.
\newblock {\em Journal of Atomic, Molecular, and Optical Physics}, 2010:128473,
  2010.

\bibitem{Monti2010JPBp205203}
J.~M. Monti, O.~A. Foj\'{o}n, J.~Hanssen, and R.~D. Rivarola.
\newblock Influence of the dynamic screening on single-electron ionization of
  multi-electron atoms.
\newblock {\em Journal of Physics B: Atomic Molecular and Optical Physics},
  43:205203, 2010.

\bibitem{Belkic1997JPBp1731}
D~Belki{\'c}.
\newblock Electron detachment from the negative hydrogen ion by proton impact.
\newblock {\em Journal of Physics B: Atomic Molecular and Optical Physics},
  30:1731--1745, 1997.

\bibitem{Monti2009JPBp195201}
J.~M. Monti, O.~A. Foj\'{o}n, J.~Hanssen, and R.~D. Rivarola.
\newblock Ionization of helium targets by proton impact: a four-body distorted
  wave-eikonal initial state model and electron dynamic correlation.
\newblock {\em Journal of Physics B: Atomic Molecular and Optical Physics},
  42:195201, 2009.

\bibitem{Monti2013PSp14031}
J.~M. Monti, J.~Fiol, D.~Fregenal, P.~D. Fainstein, R.~D. Rivarola, W.~Wolff,
  E.~Horsdal, G.~Bernardi, and S.~Su\'{a}rez.
\newblock Experimental and theoretical results on electron emission in
  collisions between partially dressed ions with {H}e targets.
\newblock {\em Physica Scripta}, T156:014031, 2013.

\bibitem{Olson1977PRAp531}
R.~E. Olson and A.~Salop.
\newblock Charge-transfer and impact-ionization cross sections for fully and
  partially stripped positive ions colliding with atomic hydrogen.
\newblock {\em Physical Review A}, 16:531--541, 1977.

\bibitem{Reinhol1986PRAp3859}
C.~O. Reinhold and C.~A. Falc\'{o}n.
\newblock Classical ionization and charge-transfer cross sections for {H}+ +
  {H}e and {H}$^{+}+${L}i$^{+}$ collisions with consideration of model
  interactions.
\newblock {\em Physical Review A}, 33:3859--3866, 1986.

\bibitem{Abrines1966PPSp861}
R.~Abrines and I.~C. Percival.
\newblock {\em Proc. Phys. Soc.}, 88:861, 1966.

\bibitem{Hardie1983JPBp1983}
D.~J.~W. Hardie and R.~E. Olson.
\newblock Charge transfer and ionisation processes involving multiply charged
  ions in collision with atomic hydrogen.
\newblock {\em Journal of Physics B: Atomic Molecular and Optical Physics},
  16(11):1983--1996, 1983.

\bibitem{Fiol2000JPBp5343}
J.~Fiol, C.~Courbin, V.~D. Rodr\'{\i}guez, and R.~O. Barrachina.
\newblock Classical description of threshold effects in ion-atom ionization
  collisions.
\newblock {\em Journal of Physics B: Atomic Molecular and Optical Physics},
  33(23):5343--5355, 2000.

\bibitem{Fiol2002JPBp1173}
J.~Fiol and R.~E. Olson.
\newblock Three-body dynamics in the ionization of hydrogen by positron impact.
\newblock {\em Journal of Physics B: Atomic Molecular and Optical Physics},
  35(5):1173--1184, 2002.

\bibitem{Bruch1975PRAp1808}
R.~Bruch, G.~Paul, J.~Andr\"{a}, and Lester Lipsky.
\newblock Autoionization of foil-excited states in {L}i i and {L}i ii.
\newblock {\em Physical Review A}, 12:1808--1824, 1975.

\end{thebibliography}

\end{document}